\documentclass[12pt]{article}
\usepackage{sc3conf}
\usepackage{amsfonts}
\usepackage{epsfig}
\usepackage[dvips]{color}

\begin{document}
\raggedbottom

\title{Interplay of correlations and fluctuations in Au+Au
collisions  at RHIC}
\authors{Mikhail L. ~Kopytine\adref{1} for the STAR Collaboration}
\addresses{\1ad Department of Physics, Kent State University, USA}
\maketitle
\begin{abstract}
Dynamic fluctuations in the local density of non-identified hadron
tracks reconstructed in the STAR TPC are studied using the discrete wavelet
transform power spectrum technique which involves mixed event reference
sample comparison. The two-dimensional event-by-event analysis is performed
in pseudo-rapidity $\eta$  and azimuthal angle $\phi$.
HIJING simulations indicate that jets and mini-jets
 result in signals, visible without high $p_T$ selection,
when the dynamic texture analysis is applied.
Scanning a broad range of event multiplicities, we study the dependence
of the signals on the initial conditions.
Event structures are studied separately with
positive and negative tracks, as well as both charges.
A change of regime is observed in AuAu collisions at $\sqrt{S_{NN}}=130$ GeV
 as event multiplicity is increased:
a long range $\eta$ correlation 
(or suppressed fluctuation vis-a-vis mixed events)
is seen in same charge data.
This effect is qualitatively similar to one of the predicted manifestations
of the Color Glass Condensate.
\end{abstract}

\section{Introduction}

Bulk properties of strongly interacting matter under extreme conditions
are the focus of the on-going RHIC program.
Deconfinement and chiral symmetry restoration\cite{Meyer-Ortmanns:1996ea}
are expected to take place
in collisions of ultra-relativistic nuclei.
Because these phase transitions are multiparticle phenomena, a
promising, albeit challenging, approach is the 
study of dynamics of large groups of final state particles. 
The dynamics
shows itself in the correlations and fluctuations (texture) on a variety
of distance scales in momentum space.

The multi-resolution dynamic texture approach 
(applied for the first time\cite{NA44} at SPS)
uses discrete wavelet
transform \cite{DWT}(DWT) to extract such information.
At present stage, the information is extracted in a comprehensive way, 
without any built-in assumptions or filters.
Mixed events are used as a reference for comparison in search for
dynamic effects. 
Event generators are used to ``train intuition'' in recognizing 
manifestations of familiar physics (such as elliptic flow or jets)
in the analysis output, as well as to quantify sensitivity to the 
effects yet unidentified, such as critical fluctuations or clustering 
of new phase at hadronization.

\section{The STAR experiment}
STAR Time Projection Chamber\cite{STAR_TPC}(TPC), 
mounted inside a solenoidal magnet,
  tracks charged particles within a large acceptance ($|\eta|<1.3$,
$0<\phi<2\pi$) and is  well suited for event-by-event physics and
in-depth studies of event structure.
The data being reported are obtained during the 
first ($\sqrt{S_{NN}}=130$ GeV) year
of RHIC operation. 
The minimum bias trigger discriminates on 
a neutral spectator signal in the Zero
Degree Calorimeters\cite{ZDC}.
By adding a requirement of high charged multiplicity within $|\eta|<1$ 
from
the scintillating Central Trigger Barrel, one obtains the central trigger.
Vertex reconstruction is based on the TPC tracking.
Only high quality tracks found to pass within 3 cm of the event vertex are 
accepted for the texture analysis.

\section{Dynamic texture analysis procedure}
\label{analysis}
Discrete wavelets are a set of functions, each having a proper width,
or scale, and a proper location so that the function differs from 0
only within that width and around that location.
The set of possible scales and locations is discrete.
The DWT transforms the collision event in pseudo-rapidity $\eta$ and azimuthal
angle $\phi$  into a set of two-dimensional functions.
The basis functions are defined
in  the ($\eta$, $\phi$) space and are
 orthogonal with respect to scale and location.
We accumulate texture information  
by averaging the power spectra of many events.

\begin{figure}
\epsfxsize=10cm
\epsfbox{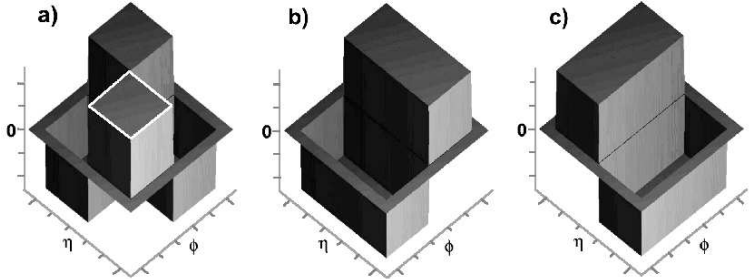}
\caption{
Haar wavelet basis in two dimensions. 
The three modes of directional sensitivity are:
a) diagonal b) azimuthal c) pseudo-rapidity. 
For the finest scale used, the white rectangle drawn ``on top'' of the 
function 
in panel a) would correspond to the smallest acceptance bin (pixel).
Every subsequent coarser scale is obtained by expanding
the functions of the previous scale by a 
factor of 2 in both dimensions. (Reproduced from {\protect\cite{NA44}}).
}
\label{haar}
\end{figure}
The simplest DWT basis is the Haar wavelet, built upon the \emph{scaling 
function}
$g(x) = 1$ for $0\le x<1$ and 0 otherwise.
The function
\begin{equation}
f(x) =     \{ +1 \mbox{\ for\ } 0\le x<\frac{1}{2}; 
                 -1 \mbox{\ for\ }  \frac{1}{2}\le x<1;
                  0 \mbox{\ otherwise}
              \}
\end{equation}
is the wavelet function.

The experimental acceptance in $\eta$,$\phi$, and $p_T$ 
(($|\eta|<1$, $0<\phi<2\pi$)) is
partitioned into bins. The $\eta$-$\phi$ partitions are of equal size,
whereas in $p_T$, the binning is exponential when more than one
$p_T$ bin is used.
In each bin, the number of 
reconstructed tracks satisfying the quality cuts is counted.

The scaling function of the Haar basis in
two dimensions (2D)
$G(\phi,\eta) = g(\phi)g(\eta)$
is just a bin's acceptance (modulo units).
The wavelet functions $F^{\lambda}$ 
(where the mode of directional sensitivity $\lambda$ can be 
azimuthal $\phi$, pseudo-rapidity $\eta$, or diagonal $\phi\eta$)
are
\begin{equation}
F^{\phi\eta}=f(\phi)f(\eta),\ \
F^\phi=f(\phi)g(\eta), \ \
F^\eta=g(\phi)f(\eta).
\end{equation}
We  set up a two dimensional (2D) wavelet basis:
\begin{equation}
F^{\lambda}_{m,i,j}(\phi,\eta) =
 2^{m}F^{\lambda}(2^{m}\phi-i,2^{m}\eta-j),
\label{wavelet_2D}
\end{equation}
where $m$ is 
the integer
scale fineness index, and
$i$ and $j$ index the positions of bin centers in 
$\phi$ and $\eta$.
Then, $F^\lambda_{m,i,j}$ with integer $m$, $i$, and $j$ are known
\cite{DWT}
to form a complete orthonormal basis in the space
of all \emph{measurable functions} defined on the continuum of real
numbers $L^2({\mathbb{R}})$.
We construct $G_{m,i,j}(\phi,\eta)$ analogously to Eq.\ref{wavelet_2D}.

Fig. \ref{haar} shows the wavelet basis functions $F$ in two dimensions.
At first glance it might seem surprising that, unlike the 1D case, both $f$ and
$g$ enter the wavelet basis in 2D.
Fig. \ref{haar} clarifies this: in order to fully encode an arbitrary
shape of a measurable 2D function, one considers it as an addition of a
change along $\phi$ ($f(\phi)g(\eta)$, panel (b)), 
a change along $\eta$ ($g(\phi)f(\eta)$, panel (c)), and
a saddle-point pattern ($f(\phi)f(\eta)$, panel (a)), 
added with appropriate weight (positive, negative or zero), for a variety
of scales.
The finest scale available is limited by the 
two track resolution, and,
due to the needs of event mixing, by the number of available events.
The coarser scales correspond to successively re-binning the track 
distribution.
The analysis is best visualized by considering the scaling function
$G_{m,i,j}(\phi,\eta)$ as binning the track distribution 
$\rho(\phi,\eta)$
in bins $i$,$j$
of fineness $m$, while the set of wavelet functions 
$F^{\lambda}_{m,i,j}(\phi,\eta)$ (or, to be exact, the wavelet expansion
 coefficients $\langle \rho, F^{\lambda}_{m,i,j}\rangle$)
gives the difference distribution between the data binned with given 
coarseness and that with binning one step finer.
We use WAILI\cite{WAILI} software to obtain the wavelet expansions.

In two dimensions, 
it is informative to present the three modes of a power
spectrum with different directions of sensitivity
$P^{\phi\eta}(m)$, $P^\phi(m)$, $P^\eta(m)$
separately.
We define the {\bf power spectrum} as
\begin{equation}
P^\lambda(m) =
\frac{1}{2^{2m}}\sum_{i,j}\langle \rho,F^\lambda_{m,i,j}\rangle^2 ,
\label{eq:P_m}
\end{equation}
where the denominator gives the meaning of spectral density
to the observable.
So defined, the $P^\lambda(m)$ of a random white noise field is
independent of $m$.
However, for physical events one finds  $P^\lambda(m)$ to be dependent
on $m$ due to the presence of {\bf static texture} features such as
acceptance asymmetries and imperfections (albeit minor in STAR),
and non-uniformity of the $\,dN/\,d\eta$ shape.
In order to extract the {\bf dynamic} signal, we use 
$P^\lambda(m)_{true}-P^\lambda(m)_{mix}$
where the latter denotes power spectrum obtained from the {\bf mixed events}.
The mixed events are composed of the ($\eta,\phi$) pixels of true events,
so that a pixel is an acceptance element of the finest scale used
in the analysis, and in no mixed event is there more than one pixel from 
any given true event.
The minimum granularity used in the analysis is $16\times16$ pixels.
\footnote{For a quick reference, here are the scales in $\eta$. Scale 1:
$\Delta\eta=1$; scale 2: $\Delta\eta=1/2$; scale 3: $\Delta\eta=1/4$ 
and so on.}

Systematic errors can be induced on
 $P^\lambda(m)_{true}-P^\lambda(m)_{mix}$ by the process of event
mixing.
For example, in events with different vertex position along the beam
axis, same values of $\eta$ may correspond to different parts of the TPC
with different tracking efficiency.
That will fake a dynamic texture effect in $\eta$.
In order to minimize such errors,
events are classified into {\bf event classes} with similar
multiplicity and vertex position.
Event mixing is done and  $P^\lambda(m)_{true}-P^\lambda(m)_{mix}$ is 
constructed within such classes.
Only events with $z$ vertex lying on the beam axis within 30 cm
from the center of the chamber are accepted for analysis.
To form event classes, this interval is further subdivided into five bins.
We also avoid mixing of events with largely different multiplicity.
Therefore, another dimension of the event class definition is that of
the multiplicity of quality tracks in the TPC.
For central trigger events, the multiplicity range of an event
class is typically 25.
For small multiplicity events taken with the minimum bias trigger, 
the multiplicity range per event class is 10 for multiplicities below 
90, and 20 in the range between 90 and 150.
For larger multiplicities in the minimum bias sample, this multiplicity
range was taken to be 50.
Events with less than 15 good quality tracks in the fiducial acceptance
are ignored.

\begin{figure}[ht]
\epsfxsize=11cm
\centerline{\epsfbox{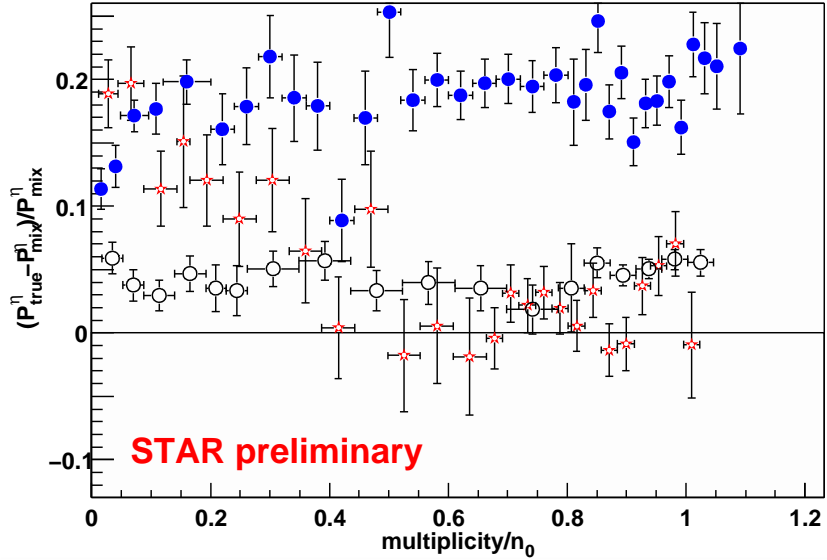}}
\caption{$(P^{\eta}_{true}-P^{\eta}_{mix})/P^{\eta}_{mix}$ 
(scale 1)
in Au Au collisions
at $\sqrt{s}=130$ GeV as a function of normalized multiplicity for 
{\bf all charged}
particles.
\textcolor{red}{$\star$} -- STAR;
\textcolor{blue}{$\bullet$} -- regular HIJING;
\textcolor{black}{$\circ$} -- HIJING without jets.
}
\label{vs_mult_C}
\end{figure}
\begin{figure}[ht]
\epsfxsize=11cm
\centerline{\epsfbox{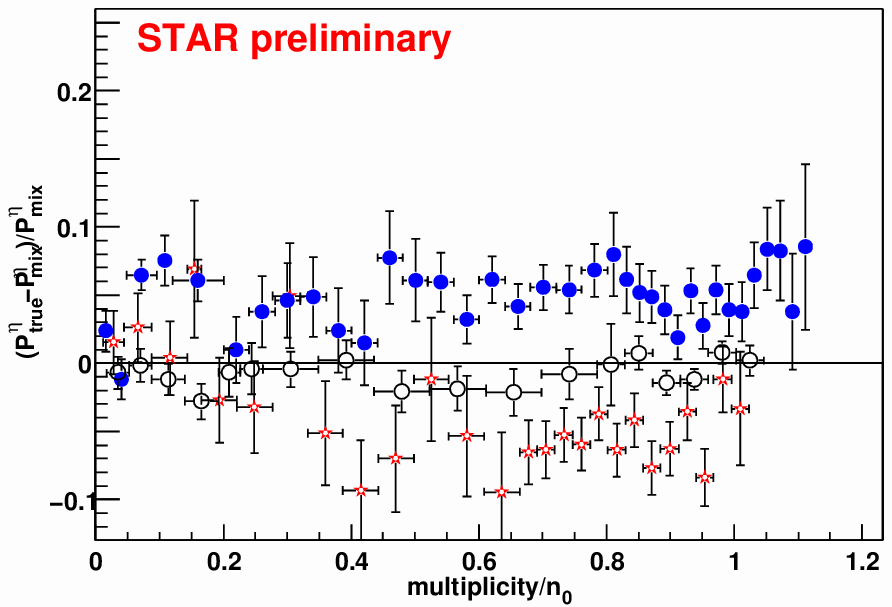}}
\caption{$(P^{\eta}_{true}-P^{\eta}_{mix})/P^{\eta}_{mix}$ 
(scale 1) in Au+Au collisions
at $\sqrt{s}=130$ GeV as a function of normalized multiplicity for 
{\bf positively} charged particles.
\textcolor{red}{$\star$} -- STAR;
\textcolor{blue}{$\bullet$} -- regular HIJING;
\textcolor{black}{$\circ$} -- HIJING without jets. }
\label{vs_mult_P}
\end{figure}
\begin{figure}[ht]
\epsfxsize=11cm
\centerline{\epsfbox{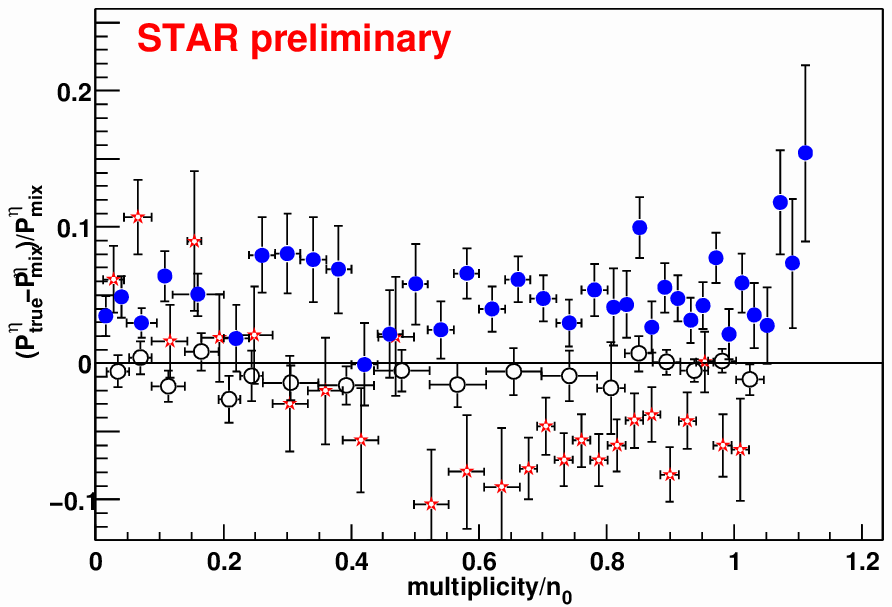}}
\caption{$(P^{\eta}_{true}-P^{\eta}_{mix})/P^{\eta}_{mix}$ 
(scale 1)
in Au+Au collisions
at $\sqrt{s}=130 GeV$ as a function of normalized multiplicity for 
{\bf negatively} charged particles.
\textcolor{red}{$\star$} -- STAR;
\textcolor{blue}{$\bullet$} -- regular HIJING;
\textcolor{black}{$\circ$} -- HIJING without jets.}
\label{vs_mult_N}
\end{figure}

\section{Dynamic textures in the STAR data}
\label{STAR_data}
Elliptic flow is a prominent large scale dynamic texture effect
already well measured at RHIC\cite{v2_RHIC}.
The DWT approach localizes elliptic flow on scales 2 and, to some degree, 3
of the $P^{\phi}_{true}-P^{\phi}_{mix}$.
In this report, we ignore flow and concentrate on the $\eta$
observables.

Multiplicity scans 
reveal dependence of the signals on the initial
conditions: number of participants and binary collisions and
the energy density and size of the interacting system. 
Such scans as a function of normalized multiplicity
\footnote{To form the normalized multiplicity, the multiplicity $n$ 
of quality tracks is divided by $n_0$, where $n_0$ is such that
99\% of minimum bias events have multiplicity of quality tracks less than 
$n_0$.
The normalized multiplicity depends weakly on the experimental definition
of minimum bias and on the track quality cuts.}
are shown in 
Fig. \ref{vs_mult_C}, \ref{vs_mult_P}, and \ref{vs_mult_N}.
The figures show the scale 1 (large scale) observables for tracks of
both charges, and for the positive only and the negative only, respectively. 
The horizontal error bars on the points are formed by multiplicity 
boundaries of event classes used in the analysis or, when a coarser re-binning
of the multiplicity is done on the analyzed data
to enhance the presentation, reflect that re-binning.
The event mixing is always confined to the specific multiplicity and 
$z$-vertex position class, described in Section \ref{analysis}.

A comparison
\footnote{
In the event generator data in the figures, 
no GEANT and no response simulation is done.
Instead, only stable charged particles ($e$,$\mu$,$\pi$,$K$,$p$) and their
antiparticles from the generator output are considered, provided that
they fit into the STAR TPC fiducial $\eta$ acceptance $|\eta| \le 1$.
Momentum resolution and $p_T$ acceptance are not simulated.
}
with HIJING\cite{HIJING} is done 
in order to understand the effects of energy-momentum conservation,
jets and resonances on our measurements.
For the regular HIJING and HIJING without jets, the
multiplicity dependence is weak
or absent.
This is understandable given the nature of HIJING as a super-position of
binary nucleon-nucleon collisions with no collective effects.
The regular HIJING and HIJING without jets show different magnitude of the
signal.
This underscores the role of jets in creating local density fluctuations
(positive dynamic texture) in HIJING.
Notice that this result is obtained without imposing a high $p_T$ cut.
The experimental data are close to regular HIJING for peripheral events,
but their texture becomes suppressed as multiplicity grows.

To understand the nature of this suppression, we look at same charge particles
(Fig. \ref{vs_mult_P} and \ref{vs_mult_N}).
We see that the STAR data are not really intermediate between the
regular HIJING and HIJING without jets, but display a more complex behavior. 
This is seen from the fact that as the multiplicity grows,
the dynamic texture becomes negative and
stays roughly constant past a transition region which occupies about a 
half of the multiplicity range, whereas the ``no jets'' prediction
points are consistent with 0 for all multiplicities.
How to explain the negative dynamic texture? Recall the definition of the
observable (Section \ref{analysis}).
At scale 1, we are looking at the fluctuation in same-charge track
occupancy between two ``pixels'', each one unit of $\eta$ long, located in
forward/backward hemispheres around mid-rapidity.
The negative $P^{\eta}_{true}-P^{\eta}_{mix}$ means that in the mixed events,
this occupancy fluctuation is stronger, i.e. a {\bf correlation} takes place
in the real events.
\footnote{This looked very counter-intuitive to us at first -- indeed, 
the events are composed out of positive and negative charge sub-events.
If the positive sub-event shows reduced (negative)
fluctuations, and so does the negative
sub-event,  how can the fluctuation measure for the combined event be 
above zero ?
To ensure that the ``balance'' is kept,
we construct and measure 0 from
$\rho$ and $\rho_{mix}$ for positive and negative tracks.
In the following, the $\langle\rho|F\rangle$ notation denotes 
 coefficients in the expansion of an individual event track density
$\rho$ into wavelet basis $F$. For simplicity, we omit the indices.
We sum over locations but not over the scale index.
Averaging over many events is assumed everywhere.
\begin{eqnarray}
{\langle \rho^+ +\rho^- | F\rangle}^2 -
{\langle \rho^+_{mix} + \rho^-_{mix}|F\rangle}^2 =\nonumber\\
 {\langle \rho^+|F\rangle}^2 + {\langle \rho^-|F\rangle}^2 
+ 2\langle \rho^+|F\rangle \langle \rho^-|F\rangle  
-{\langle \rho^+_{mix}|F\rangle}^2 - {\langle \rho^-_{mix}|F\rangle}^2
- 2\langle \rho^+_{mix}|F\rangle \langle \rho^-_{mix}|F\rangle
\label{Eq:zero}
\end{eqnarray}
We measure independently
${\langle \rho^+|F\rangle}^2 - {\langle \rho^+_{mix}|F\rangle}^2$ 
(Fig.\ref{vs_mult_P}, modulo normalization) and
${\langle \rho^-|F\rangle}^2 - {\langle \rho^-_{mix}|F\rangle}^2$
(Fig.\ref{vs_mult_N}, modulo normalization).
In order to test that Eq.\ref{Eq:zero} holds, we need to obtain
the correlation terms between the opposite charges
$\langle \rho^+|F\rangle \langle \rho^-|F\rangle$
and
$\langle \rho^+_{mix}|F\rangle \langle \rho^-_{mix}|F\rangle$.
This can be done by comparing
power spectra of images filled with equal weights for positive
and negative particles, and of those where negative particles are
entered with a negative weight:
\begin{equation}
{\langle \rho^+ +\rho^-|F\rangle}^2 -
{\langle \rho^+ - \rho^-|F\rangle}^2 =
4\langle \rho^+|F\rangle \langle \rho^-|F\rangle,
\end{equation}
which is true for $\rho_{mix}$ as well.
With this input, validity of Eq.\ref{Eq:zero} has been established and the
``paradox'' presented above resolved: it is the large scale
correlation between positive and negative charges that accounts for 
the ``extra'' texture when the positive and negative sub-events are combined.}

The tracking conditions in the chamber are different at high and low
multiplicities, and the effect of that on the measurements in question needs
to be understood. 
We do this by processing HIJING events through full GEANT and response
simulation, and applying the actual experimental reconstruction to those
events.
Such processing (done for a smaller number of high multiplicity HIJING 
events than is shown
in the figures) reveals no significant difference for the reported observable.

Thus,
we are looking at an interplay of fluctuation and correlation effects, 
neither of which is trivial, as a
function of initial conditions.
The signals become weaker for finer scales.
At this point one can only speculate about the nature of the
correlation effect in central events, 
but the HIJING simulation indicates that jet
quenching (or even total disappearance of jets) does not account for
the {\bf negative} $P_{true}-P_{mix}$, 
even though it may be a prerequisite for its observation, given
that jets work to create a {\bf positive} texture.
What about Bose-Einstein correlation?
In the simulations, its effect can not be introduced into the 
multiparticle distributions 
event-by-event from  the first principles, and in HIJING it is not
considered at all.
Low values of  $Q_{inv}$ ($<50$ MeV\cite{STAR_HBT}), typical for both
Bose-Einstein and Coulomb effects, make it unlikely for these effects
to be responsible for correlations with characteristic 
$\Delta \eta \approx 1$.

Longitudinal expansion maps rapidity differences onto time differences;
the features characterized by larger rapidity differences must be formed
early.
In the Color Glass Condensate picture, multiplicity fluctuations in
different rapidity windows are predicted to be correlated for large 
($1/\alpha_{S}$) rapidity intervals\cite{Kovchegov}.
This is seen as a consequence of the classical coherence of the gluon field.
Such or a qualitatively similar effect is indeed required to explain the data.

\section{Conclusions}
The STAR measurements of AuAu data reveal a non-trivial picture of 
non-statistical correlations and fluctuations which is qualitatively 
different for peripheral and central collisions. A possible interpretation
of the central data is suppression of jets, combined with particle emission
correlated over the length of the order of a unit of rapidity.

\end{document}